# Machine Learning for Identifying Grain Boundaries in Scanning Electron Microscopy (SEM) Images of Nanoparticle Superlattices

Aanish Paruchuri [1], Carl Thrasher[2], A. J. Hart[3], Robert Macfarlane[2] Arthi Jayaraman [4,5,6] *

1. Master of Science in Data Science Program, University of Delaware, Newark DE 19713
2. Department of Materials Science and Engineering, Massachusetts Institute of Technology, Cambridge, MA, 02139
3. Department of Mechanical Engineering, Massachusetts Institute of Technology, Cambridge, MA, 02139
4. Department of Chemical and Biomolecular Engineering, 150 Academy St, University of Delaware, Newark DE 19713
5. Department of Materials Science and Engineering, University of Delaware, Newark DE 19713
6. Data Science Institute, University of Delaware, Newark DE, 19713

**\* Corresponding author**

arthij@udel.edu

**Abstract**
Nanoparticle superlattices consisting of ordered arrangements of nanoparticles exhibit unique optical, magnetic, and electronic properties arising from nanoparticle characteristics as well as their collective behaviors. Understanding how processing conditions influence the nanoscale arrangement and microstructure is critical for engineering materials with desired macroscopic properties. Microstructural features such as grain boundaries, lattice defects, and pores significantly affect these properties but are challenging to quantify using traditional manual analyses as they are labor-intensive and prone to errors. In this work, we present a machine learning workflow for automating grain segmentation in scanning electron microscopy (SEM) images of nanoparticle superlattices. This workflow integrates signal processing techniques, such as Radon transforms, with unsupervised learning methods like agglomerative hierarchical clustering to identify and segment grains without requiring manually annotated data. In the workflow we transform the raw pixel data into explainable numerical representation of superlattice orientations for clustering. Benchmarking results demonstrate the workflow's robustness against noisy images and edge cases, with a processing speed of four images per minute on standard computational hardware. This efficiency makes the workflow scalable to large datasets and makes it a valuable tool for integrating data-driven models into decision-making processes for material design and analysis. For example, one can use this workflow to quantify grain size distributions at varying processing conditions like temperature and pressure and using that knowledge adjust processing conditions to achieve desired superlattice orientations and grain sizes.

# I. Introduction

Nanoparticle superlattices are ordered arrangements of nanoparticle building blocks that produce interesting optical, magnetic, and electronic properties.[1] These material properties are achieved from the individual properties of the nanoparticles as well as the collective behaviors that arise from their relative spatial arrangement.[2] To construct devices with functional nanoparticle superlattices there is a need for improved understanding of how processing of these materials affects their nanoscale arrangement and microstructure.[3], [4] Microstructural features such as lattice defects, pores, and grain boundaries are hypothesized to have large effects on the macroscopic properties of nanoparticle superlattices, similar to their role in atomic crystalline materials.[5] While these structural features can be observed visually in nanoparticle superlattices via scanning electron microscopy (SEM), these features can be challenging to *quantify* manually. For example, one report has successfully demonstrated the assembly of polycrystalline nanoparticle superlattice materials with rudimentary control of grain size, but the methods used to characterize the grains involved tedious manual identification and mapping of each grain in the microscopy images.[6] Such challenges are in contrast to the typical grain analysis in atomic materials for which electron backscatter diffraction (EBSD) techniques allow the direct mapping of grains to capture size, orientation, and shape distribution.[7] While it is possible to perform Williamson-Hall analysis to get a measure of nanoparticle superlattice grain size using small-angle X-ray scattering (SAXS) techniques, this necessitates synchrotron beamline access, yields only an average grain size (instead of a distribution of the grain sizes), and is vulnerable to noise (signal ambiguity stemming from instrument broadening, testing conditions, sample quality, etc.).[8], [9] Thus, a method to quickly and reliably identify grains and quantify the grain size distribution for nanoparticle superlattice cross-sections would be beneficial for engineering polycrystalline nanoparticle superlattice materials with desired structural features and properties.

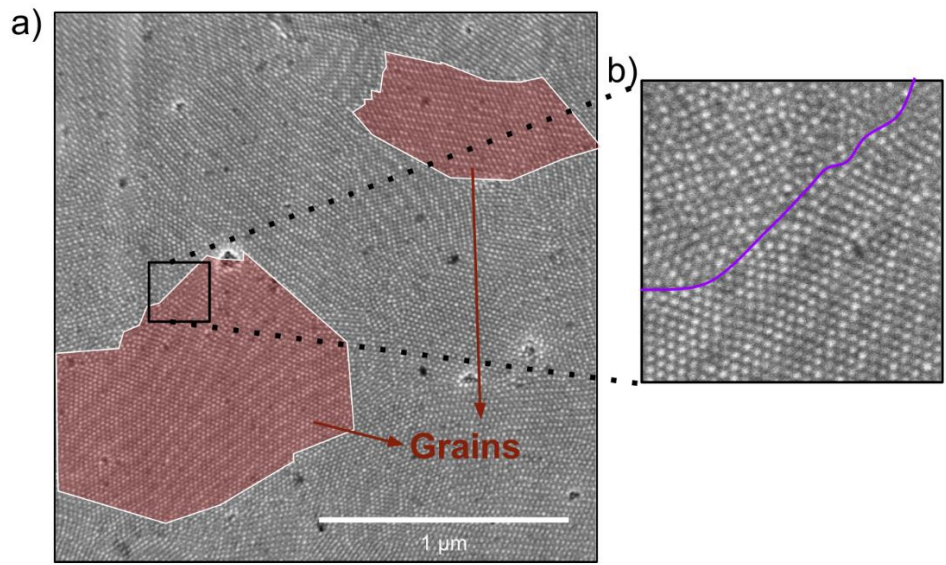

*Figure 1*: *(a) Example of an SEM image with nanoparticle superlattices; colored regions denote two distinct grains and the purple line in (b) denotes part of the grain boundary [boundaries are also shown as white outlines for the grains shown in part (a)].*

Machine learning (ML) models, specifically related to pattern recognition, can be used for identifying features of interest in SEM images[10], [11]. Specifically, image segmentation models divide an image into distinct regions or segments based on image features such as image texture, intensity, or pattern. For images capturing nanoparticle superlattices, ML-based segmentation has the potential to precisely identify a grain defined as a region of the superlattice with the same orientational arrangement of nanoparticles, boundaries between grains, and other structural features (grain shapes and sizes), thus, offering a better alternative to time-consuming and error-prone manual analysis methods. Traditionally, image segmentation has relied on image processing techniques to identify homogeneous regions of interest in images with 'homogeneity' having a predefined criterion. With the advent of deep learning, image segmentation has shifted towards learning-based approaches, where convolutional neural network (CNN) based models leverage large datasets to learn and predict image segments [12], [13]. A significant drawback of CNNs is the requirement for large, labeled training datasets [14], [15]. Although, these models, initially developed for general-purpose tasks of segmenting everyday objects like cars, people, and animals, have now been successfully adapted for microscopy image analysis through transfer learning [16], [17]. Transfer learning leverages a previously developed model that has been trained on general datasets and fine-tunes parts of the model (e.g., layers of the neural network) with new data to learn aspects specific to microscopy images. While transfer learning reduces the need for extensive datasets by repurposing pre-trained models, it still requires hundreds of accurately annotated examples of the task-specific images for effective training [18]. This poses a challenge in the current context, as manually annotating grain boundaries is labor-intensive, time-consuming, and prone to subjectivity, leading to inconsistencies in the annotations.

To address limitations of transfer learning model training, unsupervised learning offers a promising alternative by eliminating the need for annotated data. In unsupervised learning, the model identifies patterns and relationships within the data by grouping regions into clusters based on similarities and dissimilarities in the microscopy images. This clustering approach not only avoids the need for extensive manual intervention but also generalizes well across different samples, making it highly adaptable to diverse datasets and experimental conditions without requiring additional training. Furthermore, studies show that signal processing techniques, such as the Radon transform, Fourier transform, or wavelet decomposition, enhances the unsupervised learning workflow[19]. By transforming raw pixel data into a feature-rich domain, these techniques capture essential structural characteristics of the grains, such as orientation of the lattice of nanoparticles and spatial frequency of nanoparticles, which are critical for accurate segmentation. This transformation from raw pixels to feature-rich domains reduces noise and emphasizes only the meaningful variations in the data (rather than noise), significantly improving the performance of clustering algorithms.

In this paper, we build on our previous work using signal processing techniques and unsupervised machine learning for AFM image analysis [19] to develop a new robust workflow for segmenting grains of nanoparticle superlattice cross-sections from SEM images. Through the newly developed workflow, we accurately identify grain boundaries and segment individual grains in SEM images with minimal manual intervention. Additionally, the grain size distributions quantified by this workflow for SEM images obtained at varying processing conditions shows if and how processing temperature and pressure affect the microstructural features in the nanoparticle superlattices.

## II. SEM Image Datasets from Experiments

The nanoparticle superlattices in consideration in this paper are made up of nanoparticle tectons (NCTs) composed of a nanoparticle core grafted with multiple polymer chains forming a brush. The grafted polymer chains are end-functionalized with one of two chemical species that form a pair of supramolecular binding groups. In our study the pair of binding groups are diaminopyridine (DAP) and thymine (THY), which bind to each other specifically due to complementary hydrogen bonds. The polymer chains on each NCT are end-functionalized with either DAP or THY. The formation of hydrogen bonds between the NCTs' grafted polymer ends leads to self-assembly of the NCTs. The NCTs self-assemble to form single-crystalline nanoparticle superlattices with distinct crystallographic order, (that is, CsCl structure), as reported previously[20], [21], [22]. NCT crystallites can also be fused into macroscopic polycrystals through a process termed colloidal sintering, which involves the application of pressure to colloidal phase superlattices via rapid centrifugation[6]. This process produces macroscopic materials that retain nanoscale ordering.

To understand how to manipulate meso-to-microscale features (i.e., grains, pores) in these materials, we sinter a series of NCT polycrystals under varying processing conditions of temperature and pressure. We characterize the internal structure of these materials using focused ion beam (FIB) milling with FEI Helios Nanolab 600 Dual Beam System. Before milling, 30 nm of gold is deposited on samples (via AJA eBeam evaporator) to aid conduction. Cross-sectional micrographs are collected on the dual beam system with a 52° relative difference between the ion and electron beam. Samples are milled with a 6.5 nA (30 kV) Ga ion beam, followed by a cleaning cross-section milling using a 2.3 nA (30 kV) Ga ion beam. Each cross-section is imaged with an 86 pA (5 kV) electron beam using the in-lens detector without using the software's tilt correction.

The experimentally obtained SEM images are collated broadly based on sample processing conditions (i.e., temperature and pressure) and different cross-sectional regions of a single sample. The SEM renders measurements as TIFF format images, attached with metadata detailing imaging settings and length scale information. **Supporting Information Figure S1** shows examples of data points in the dataset which depict the substantial variability in magnification, contrast, and the extent of defects across the dataset. SEM images displayed a noticeable line noise pattern **Figure S1 (b, d)**, which obscures finer details. Additionally, defects within the samples are in two distinct forms where some exhibit a grain-like pattern **Figure S1.b**, while others appear as dark regions **Figure S1 (a, d)**. For effective image segmentation, grain patterns are clearly distinguishable at magnifications of approximately 50,000×. We also notice some samples being out of focus leading to blurry images. Such images are excluded from further analysis. This variability in samples is less desirable and with the help of preprocessing we standardize the contrast, magnification, and image quality. Additionally, we exclude suboptimal images to achieve accurate segmentation and grain analysis.

## III. Computational Methods for SEM Image Analyses

From a computational perspective, the data in an SEM image is a grid of points where each point holds intensity values. SEM images are like gray scale images where the intensity values are depicted as pixels; in contrast, colored images hold red, green blue color information at each pixel. The intensity variations in the SEM images in our dataset visually looks like a collection of beads where each bead represents a nanoparticle, and the beads' spatial arrangement represents superlattice pattern. Our goal is to identify grains where each grain is considered as a region of the SEM image where the beads/nanoparticles in a superlattice have a consistent orientation of the superlattice vector throughout the grain. This process of segmenting regions based on their features (i.e. superlattice orientation) is referred to as region-based image segmentation. The task at hand

– identification of grains - cannot utilize the state-of-the-art region-based image segmentation techniques (e.g., deep learning models) as they require large datasets for effective transfer learning; we have limited data (< 100 images) at hand. Consequently, as outlined in **Section I**, there is a need for unsupervised learning approaches to address the task at hand.

Unsupervised learning models such as clustering, group similar features together and dissimilar features apart. To achieve this grouping, there is a measure of distance or similarity that is required to be quantified. In the context of our experimental dataset, similarity could be characterized by tiles with consistent intensity variations that correspond to specific nanoparticle arrangements or superlattice orientations. Dissimilarity, on the other hand, arises from transitions between regions with differing superlattice orientations or distinct patterns in nanoparticle arrangements. To enable clustering based on these similarity or dissimilarity characteristics, the pixel-based SEM image data must be transformed into a feature space that effectively encodes the meaningful patterns present. One effective approach is to use the Radon transform, which captures orientational information and structural details of the superlattice. We could extract features from the Radon transform of SEM images and define a feature space. In this context, the feature space is defined by these extracted features, with each point in the space representing a specific grain orientation or structural pattern.

To effectively perform grain segmentation on SEM images we define a workflow (**Figure 2**) in which we first preprocess the SEM images, then tile it into smaller sections, apply Radon transformations on each tile, extract features from the sinogram, and lastly, cluster the features where similar tiles of a grain form a cluster. We will describe each of these steps of the workflow in more detail next.

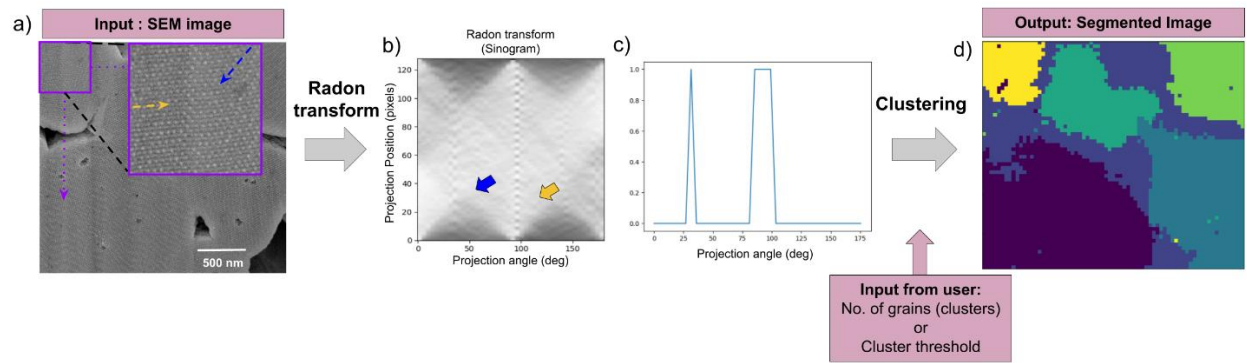

*Figure 2: Domain transform workflow. (a) Preprocessed SEM image is broken into overlapping subsections called tiles (purple box is zoomed in). (b) Tiles are converted into a sinogram using domain transforms (in our case Radon transform). The sinogram with statistics is further reduced to (c) a feature vector which is clustered to make (d) a segmented image.*

A. **SEM image pre-processing**

The SEM images obtained from the instrument are in raw TIFF image file format which often come with attached metadata. From the TIFF files the metadata is separated and the pixel information is extracted into a PNG file that is passed down to the workflow. Additionally, the samples often include defects (i.e., regions devoid of grains) that must be excluded from further analysis. To address this, defect identification is performed as a preprocessing step, where image morphological operations and image thresholding are applied to segment and mark defective regions in black. This step ensures that these defective areas are effectively removed from the analysis within the workflow. Furthermore, variations in contrast and magnification between samples can impact consistency and analytical performance. To standardize the images and enhance performance, histogram equalization is performed to ensure uniformity in contrast. Additionally, analysis is performed on samples with similar magnification to maintain uniformity in grain aspect size.

**B. Tiling**

The post-processed SEM images are divided into overlapping subsections, referred to as 'tiles'. Each tile encompasses a localized neighborhood of a pixel; tile size is a tunable parameter aimed to capture sufficient spatial context to represent grain structures effectively. Tile size is experimentally determined for the dataset [19]; in this case 128-pixel square tile size works best with SEM images of magnification 50,000X. By capturing the spatial relationships and intensity variations within these neighborhoods of the pixel, the tiles retain the essential features required to differentiate grains. Clustering algorithms applied to these tiles can then leverage this localized information to identify and distinguish unique grains within the image. In summary, tiles ensure that the analysis considers the spatial continuity and structural characteristics of the grains.

**C. Radon transform**

Although tiles represent the structural characteristics of grains visually, clustering applied directly to tiles that hold pixel intensities is insufficient to identify patterns effectively. To enhance the performance of clustering, it is beneficial to apply the Radon transform to the tiles and extract explicit superlattice orientation information.

Radon transform works by projecting the intensity values of an image along a specified set of angles, effectively converting spatial information into a domain that emphasizes orientation and line features. For a given tile, the transform integrates pixel intensity values along straight lines at various angles know as 'projection angle', generating a 'sinogram'. The sinogram is a 2D matrix that represents the variation of integrated intensity in the vertical direction and projection angle in the horizontal direction as shown in **Figure 3**.

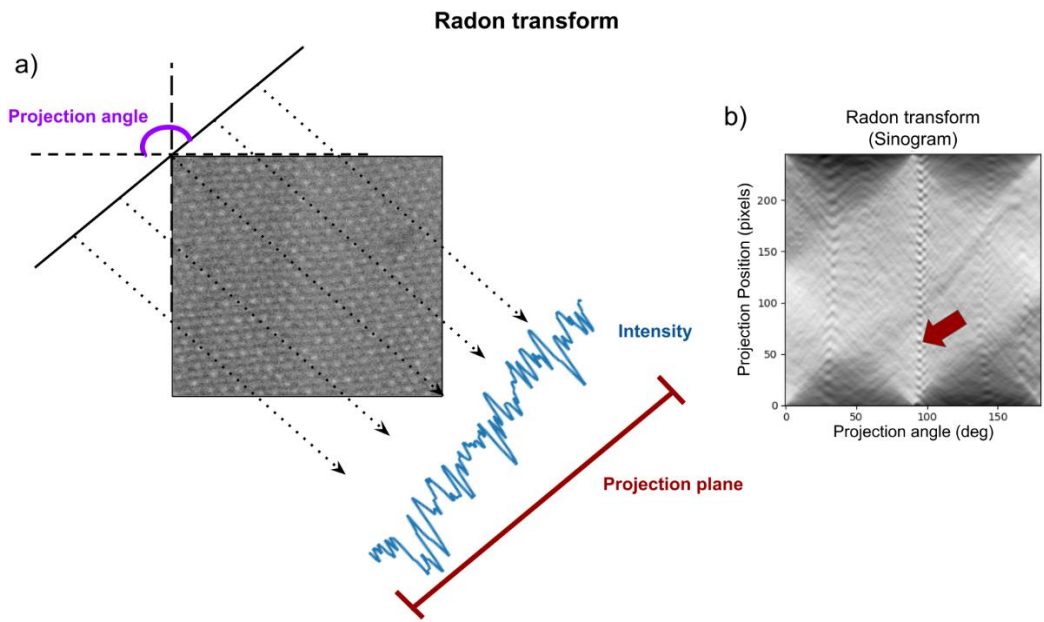

*Figure 3. Visualization of Radon transform: (a) For varying projection angles, each of the dotted rays perpendicular to the projection plane integrates the pixel values along its path in the SEM image. These integrated values are then projected onto the projection plane, as illustrated by the blue plot on the red projection plane. Stacking the projections horizontally for all projections yields (b) the sinogram.*

The sinogram from the Radon transform highlights structural patterns in the tile, such as superlattice orientations, which become explicitly visible as vertically aligned binary patterns at specific projection angles. These patterns occur because the bead-like nanoparticles are aligned with the line integral of the Radon transform at some angles and produce distinct intensity variations, forming visually distinct features in the sinogram. Consequently, such binary patterns appear at multiple projection angles, corresponding to the unique orientations of grains within the tile. These aligned projection angles contain sufficient and distinctive information to effectively identify and differentiate grain information in tiles.

To efficiently utilize this information for clustering, the sinogram is reduced to a feature vector that encodes the alignment angles. This dimensionality reduction is achieved by analyzing the variance of intensity values across all projection angles. Specifically, for each projection angle in the sinogram, the intensity variance is computed along the vertical axis, capturing the degree of alignment for that angle. A threshold is then applied to the variance values to distinguish aligned (binary 1) from non-aligned (binary 0) projection angles. This results in a fixed-length binary vector, referred to as the feature vector, where each entry corresponds to a projection angle and indicates whether a significant alignment is present. The feature vector is computationally efficient to work with due to its small size and fixed length, making it suitable for clustering algorithms. By encoding only the alignment angles, the feature vector efficiently captures essential orientation information from the sinogram while minimizing computational complexity.

### D. Clustering

In the clustering process, the feature vector derived from each tile is used to measure the similarity or dissimilarity between tiles, enabling the assignment of grain indexes. A 'grain index' is a unique numerical assigned to a tile to represent the grain to which it belongs. These indexes are determined based on the clustering model's ability to group tiles with similar orientational characteristics. The clustering model processes the feature vectors and outputs a grain index for each feature vector. This grain index is then mapped back to the corresponding tile, resulting in the assignment of a specific grain index label to each tile. By arranging these grain indexes in the same spatial order as the tiles within the original image, a grain index map is generated. This map provides a spatial representation of the segmented grains, where each unique index corresponds to a distinct grain, enabling the visualization and analysis of the grain structure across the SEM image.

The specific clustering model used can play a crucial role in the ability of the workflow to segment grains. In the following subsection we will review the portioning-based K-means model and hierarchical-based clustering model along with two different distance metrics - Euclidean and dynamic time wrapping (DWT).

### a. K-means clustering

K-means is a popular clustering model designed to partition data into a predefined number of clusters based on similarity. It starts by randomly selecting initial cluster centers, known as centroids, and iteratively adjusting the centroid positions to minimize the total distance between the centroids and the feature vectors assigned to them. Each feature vector is assigned to the cluster with the nearest centroid based on a chosen distance metric, which is recalculated after every iteration until convergence is achieved.

A commonly used distance metric in K-means is Euclidean distance, which measures the straight-line distance between points in a multi-dimensional Euclidian feature space. In the context of our application, the dimensions of this Euclidean feature space correspond to the number of projection angles in the Radon transform. The Euclidean distance naturally organizes feature vectors with similar grain orientations close to each other in this multi-dimensional space. This is because the metric effectively captures differences in the magnitude and direction of feature vectors, which are strongly related to the orientation pattern of superlattices. As a result, feature vectors from grains with similar super lattice orientations gravitate toward the same centroid during the iterative assignment process.

By iteratively adjusting centroids and reassigning feature vectors, K-means ensures that clusters become increasingly distinct. This process facilitates the formation of definitive clusters where each cluster ideally corresponds to grains with similar superlattice orientation characteristics. The final centroids serve as representations of the common characteristics of the grains within each cluster, providing a straightforward way to differentiate grains.

b. **Dynamic time wrapping**

Dynamic Time Warping (DTW) is a distance measure used in clustering that is particularly suited for analyzing patterns with misalignments or distortions. In the context of grain segmentation, superlattices with similar orientations (i.e., in the same grain) may exhibit slight shifts or distortions in their feature vectors due to imaging artifacts or local variations in nanoparticle alignment in the superlattice. DTW potentially mitigates these discrepancies by dynamically warping the feature vectors along the sequence, aligning them in a way that minimizes the cumulative distance between their corresponding elements. Unlike Euclidean distance, which rigidly compares corresponding elements of two feature vectors in a one-to-one fashion, DTW accommodates variations by enabling nonlinear alignments.

This alignment flexibility enhances K-means clustering by allowing superlattices with minor misalignments to be grouped in the same cluster, effectively preserving grain continuity. However, the primary drawback of DTW is its computational cost. Calculating the optimal alignment path of feature vectors involves constructing a cost matrix and applying dynamic programming, which has a time complexity of $O(n^2)$ for each pairwise comparison, where $n$ is the length of the feature vectors. This can significantly increase computation time when dealing with large SEM images, making DTW more resource-intensive compared to simpler metrics like Euclidean distance.

c. **Hierarchical clustering**

Hierarchical clustering is another widely used clustering method that builds a hierarchy of clusters by successively merging or dividing them based on their similarity. Unlike K-means, which requires the number of clusters to be predefined, hierarchical clustering creates a dendrogram, or tree-like structure, which represents the nested relationships among feature vectors. This allows for flexible cluster determination, as the number of clusters can be decided by cutting the dendrogram to the desired level. This feature of hierarchical clustering improves the automation effort by reducing the manual input of number of clusters beforehand.

The assessment of similarity or dissimilarity between clusters is dependent on two factors, the distance metric and the linkage type. Euclidean distance is a commonly used metric that measures similarity of feature vectors and the linkage criterion defines how distance between clusters is computed. For instance, single linkage with Euclidean distance gives the Euclidean distance between two of the closest points in the cluster whereas average linkage gives the average distance of all the points in two clusters.

In the context of grain segmentation, agglomerative hierarchical clustering (i.e., clustering by successively merging) offers the ability to capture nested relationships among feature vectors derived from the Radon transform's sinograms. Initially, each feature vector is treated as an individual cluster, and at each iteration, the two most similar clusters are merged. This iterative process continues until all feature vectors are grouped into a single cluster or until a stopping criterion is met. This hierarchical approach provides flexibility in determining the number of clusters through either predefining the number of clusters and stopping once the desired clusters are formed or a distance threshold which is a dynamic stopping criterion that stops the iterations when the distance between clusters reduces below a certain distance threshold.

The distance threshold can be experimentally determined by analyzing the distance metric between tiles containing superlattices with similar orientations and those with slight variations. This process involves calculating the pairwise distance between feature vectors representing the tiles and evaluating the range of distances that distinguish grains with subtle superlattice orientation differences from those with the same superlattice orientation. The optimal stopping criterion is identified as the threshold that maximizes the separation between these two groups. This threshold ensures that tiles within the same grain are consistently grouped together, while distinct grains are accurately segmented. Using a distance threshold improves automation by eliminating the need for predefined cluster numbers, allowing the algorithm to adapt to the data's inherent structure. Unlike K-means, hierarchical clustering does not require an initial random selection of centroids, which eliminates potential variability in results due to initialization. However, it is computationally more intensive than K-means.

### E. Post-processing the grain index map

The grain index map obtained from the clustering models is further refined through a series of post-processing steps to improve the accuracy and reliability of the predictions. Post-processing techniques are problem-specific and are tailored to address limitations inherent to the clustering models, which often fail to meet subtle performance criteria. In our case, post-processing is employed to remove noise, ensure grain continuity, and separate spatially disconnected grains into individual grains. This is achieved by sequentially applying image processing techniques such as image denoising, image morphological operations and low pass filtering, to each grain's binary map. The binary map of a grain in this case is obtained from the grain index map. These techniques refine the segmentation by eliminating noise, smoothing boundaries, and separating fragmented grains. After post-processing, the individual binary maps are collated, and unique numerical grain indexes are assigned to each segmented region yielding a post-processed grain index map.

## IV. Results

We first evaluate the performance of the clustering models—K-means and Agglomerative Hierarchical Clustering—using the two distance metrics - Euclidean and Dynamic Time Warping (DTW) as discussed in **Section III**. This evaluation aims to assess how effectively these models and metrics integrate into the overall workflow. Specifically, we focus on their ability to accurately cluster feature vectors, facilitate automation, and demonstrate reliability when handling variations in grain orientation and imaging artifacts such as line noise, defects, or blurred regions. Due to the limited availability of manually annotated data for validation, quantitative analysis is constrained.

Therefore, our evaluation emphasizes *qualitative* observations from a diverse set of test cases. We examine the models' performance in challenging scenarios, including noisy data and edge cases, to assess their robustness and practical applicability.

Additionally, we apply our workflow to calculate the grain size distribution observed in SEM images obtained from samples subjected to varying experimental conditions (temperature and pressure). The results of our analysis are presented, highlighting the observed trends and key findings. This investigation provides insights into how experimental parameters influence the grain size distribution, demonstrating the utility of our workflow for quantitative characterization.

### A. Assessment of distance metric

We find that the choice of the distance metric significantly impacts the performance of clustering models within the grain segmentation workflow. To evaluate the impact of these metrics, we compare Euclidean distance and DTW in clustering feature vectors derived from the Radon transform. Both metrics are applied within the K-means algorithm to group feature vectors corresponding to grains in SEM images. As shown in **Figure 4**, segmentations produced using the DTW metric exhibit discontinuity and result in noisy segmentation patterns. This lack of spatial coherence in the grain boundaries suggests that DTW struggles to maintain the structural integrity of grains. One plausible explanation for this failure is that DTW prioritizes minimizing the cumulative alignment cost across the sequence of the feature vector, leading to overfitting on pattern within feature vectors and ignoring projection angle. Consequently, DTW groups tiles with minor pattern misalignments into separate clusters, causing fragmented and noisy segmentations.

In contrast to DTW, the Euclidean distance metric provides more coherent and continuous grain segmentations that closely align with the actual grain boundaries observed in SEM images. However, minor noise artifacts are observed, primarily along boundary regions where feature vectors exhibit intermediate orientations. These artifacts are minimal and are effectively mitigated through post-processing techniques such as image morphological operations.

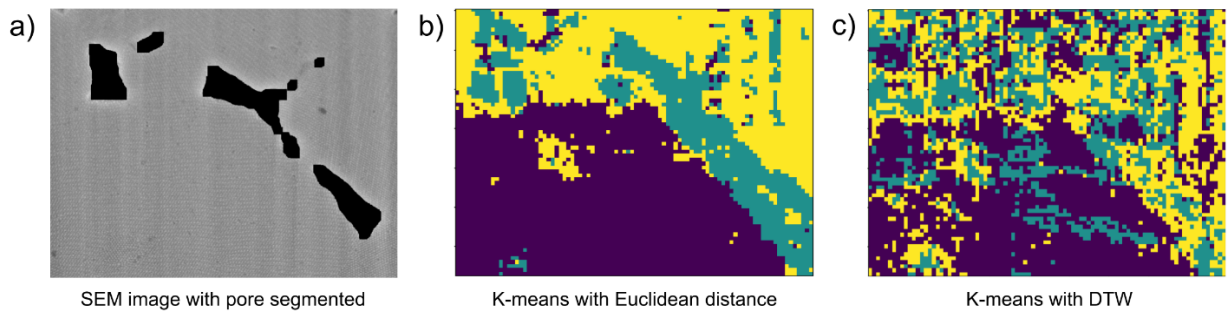

*Figure 4. Comparison of K-means and Agglomerative Clustering performance:* *The figure includes the preprocessed SEM image (a), and the segmented outputs generated using the Euclidean distance (b) and Dynamic Time Warping (DTW) (c) metrics with the K-means algorithm, configured to predict three clusters.*

The comparison between DTW and Euclidean distance becomes even more evident when examining their quantitative performance in distinguishing feature vectors. The computed distance values for DTW and Euclidean distance when applied to feature vectors representing similar and dissimilar grains shows that DTW consistently produced lower distance values for closely resembling patterns, even when the projection angles differed significantly. An example is presented in **Supporting Information Figure S2.** This characteristic of DTW is advantageous to preserve grain continuity where there is minor misalignment in projection angles but its unsuitable for grain segmentations where the angular alignment of grains is also critical for accurate clustering. DTW's flexibility in warping patterns leads to a reduced sensitivity to projection angle variations which is counterproductive. In contrast, the Euclidean distance metric demonstrates a higher sensitivity to angular alignment, effectively capturing differences in orientation of superlattices. The metric's ability to differentiate grains based on their Radon transform feature vectors ensures that clusters represent distinct grains.

**B. Assessment of clustering models**

From **Section IV.A**, we conclude that Euclidean distance is a more effective metric for use in the grain segmentation workflow. Building on this conclusion, we evaluate the performance of two clustering models, K-means and agglomerative hierarchical clustering, using Euclidean distance in our workflow. To individually assess each model's performance, we do not postprocess the output. Upon visual validation, both models demonstrate similar performance, producing well-defined grain boundaries in most cases. However, agglomerative hierarchical clustering consistently exhibits better noise reduction in the segmentation output compared to K-means. An example illustrating this difference is provided in **Figure 5**, where agglomerative hierarchical clustering achieves cleaner grain boundaries with fewer noisy artifacts in the output. Agglomerative hierarchical clustering's ability to iteratively merge clusters based on pairwise similarity enables it to maintain hierarchical relationships among feature vectors. Unlike the centroid-based assignment in K-means, which performs relatively poorly in grain boundaries with intermediate grain orientations and ambiguous alignments.

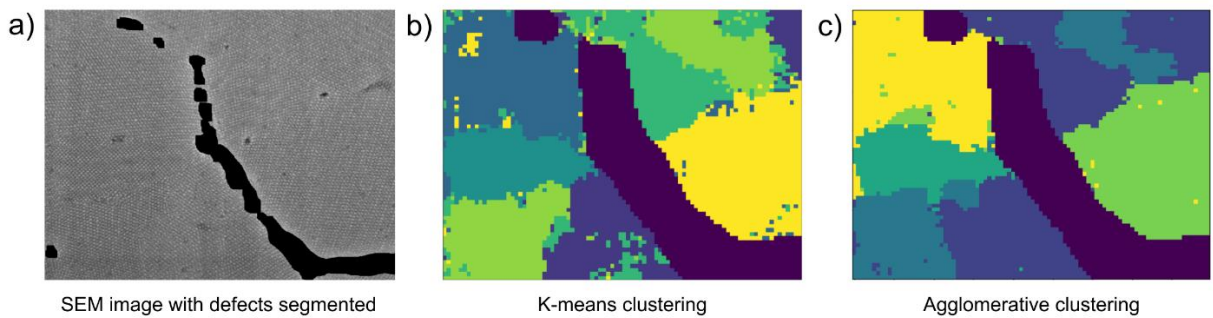

*Figure 5. Comparing performance of K-means clustering and agglomerative hierarchical clustering. (a) preprocessed SEM image and its corresponding grain segmentation with (b) K-means clustering and (c) agglomerative hierarchical clustering workflows where each model was tasked to predict 8 clusters.*

One key advantage of agglomerative hierarchical clustering is its ability to use a dynamic stopping criterion, such as a distance threshold as discussed in **Section III.D**. This dynamic stopping criterion eliminates the need to predetermine the number of clusters. This threshold-based approach has allowed for consistent performance across SEM images of a given sample, improving the automation of the segmentation workflow. In contrast, K-means requires the number of clusters to be defined input beforehand, which may necessitate trial and error to achieve optimal results leading to high computational cost.

In summary, while both K-means and agglomerative hierarchical clustering models are effective, agglomerative hierarchical clustering provides better segmentation quality when analyzed without post processing. The dynamic stopping criterion based on threshold-based mechanism enhances the unsupervised nature of the solution, making it a preferred choice for grain segmentation tasks.

**C. Evaluation of segmentation workflow performance**

The workflow produces a post-processed grain index map that spatially identifies the locations of grains within the SEM image. This map serves as the foundation for further analysis, enabling the extraction of critical grain characteristics such as size, super lattice orientation, and distribution. The primary objective of the workflow is to generate accurate grain segmentations, which can then be used to quantify grain properties.

**Figure 6** illustrates the results of grain segmentation achieved using Euclidean distance combined with agglomerative hierarchical clustering, with a tile size of 128 pixels; this is what we mean by 'optimal workflow' in the figure caption. To visualize the segmentation results, we color-code the grain indexes to produce a "predicted segment map". From **Figure 6**, we infer that the predicted segment map demonstrates the workflow's ability to accurately identify the bulk regions of grains consistently across samples highlighting the workflow's generalizability. Additionally, the workflow is robust to line noise from the input SEM image and the agglomerative clustering distance threshold accurately identifies the correct number of grains.

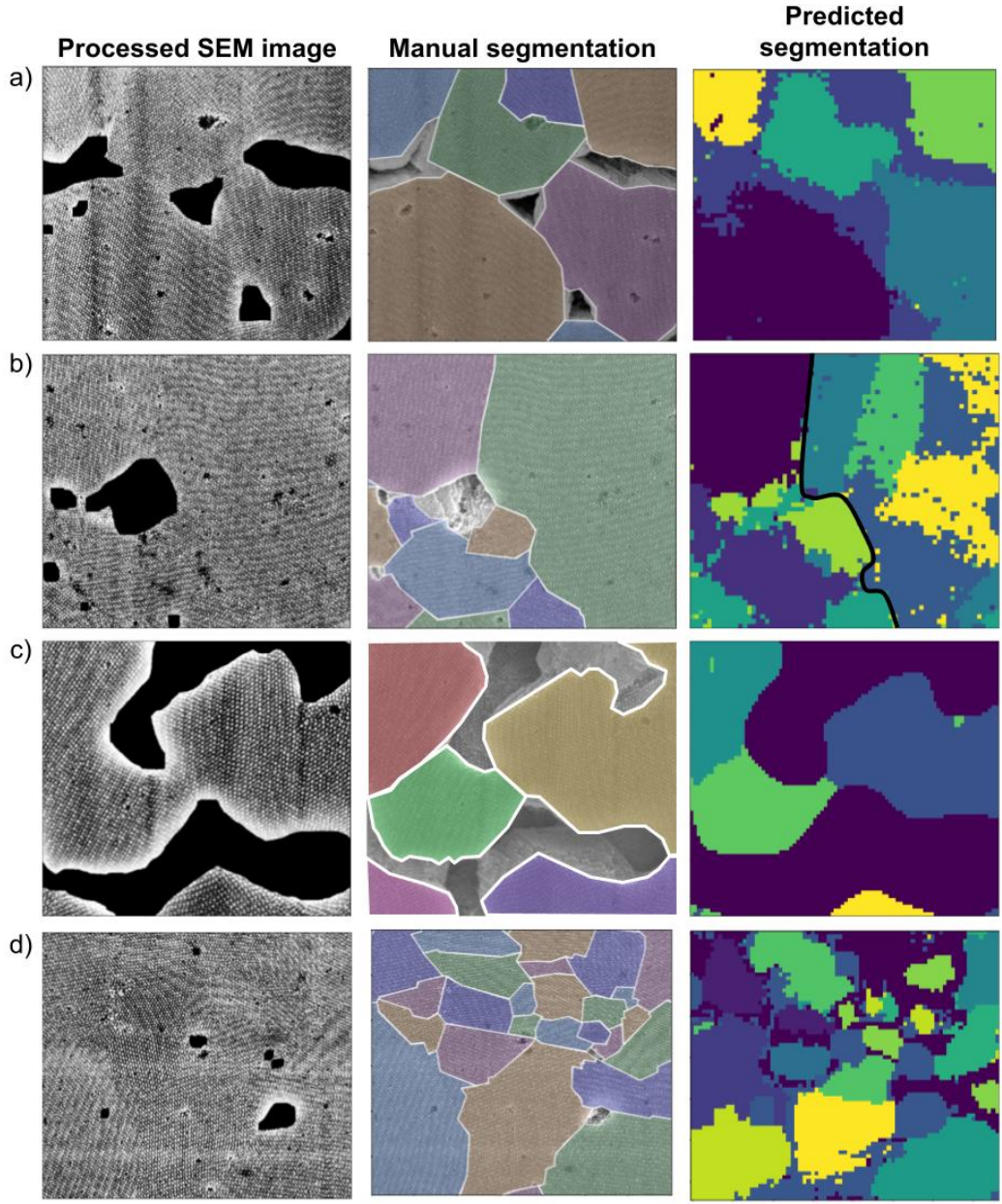

*Figure 6.* *Comparison of predicted segmentation results from post-processed SEM images using the optimal workflow against corresponding manual annotations.*

There are certain characteristics of the workflow that warrant closer attention. One notable observation is the tendency for grains with similar orientations but spatially distinct locations to be assigned the same grain index. This occurs due to the similarity in feature vectors derived from these grains, which the clustering algorithm groups together. To address this, post-processing is used to identify large, disconnected regions to segment into individual grains while preserving their size and spatial integrity. An example of this is illustrated in **Figures 6b, d,** where grains are initially assigned the same index. Additionally, there are instances in **Figures 6a, b** where defects and grains are assigned the same index. In such cases, during postprocessing we subtract the positions of defects to distinguish defects from grains unless the defect exhibits significant connectivity with the grain sharing the same index. This ensures that the grain index map remains representative of the true microstructural features. These nuances highlight the robustness of the workflow in addressing complex scenarios while maintaining the flexibility for refinement during post-processing.

We identified two notable cases in the segmentation results we wish to discuss in additional detail. The first case is where a single grain is fragmented into multiple smaller grains, as shown in **Figure 6b** (grain manually segmented in green). Upon closer examination, we observed that the grain in question exhibited significant variation in orientation within its boundaries. This behavior aligns with the workflow's design, which prioritizes detecting considerable changes in grain orientation. While this segmentation characteristic is intentional, it highlights the workflow's sensitivity to orientation gradients within grains, also referred to as paracrystalline distortion[6], [23]. The second case is where the segmentation of small grains becomes noisy, as illustrated in **Figure 6d**. In this scenario, the grain size is smaller than the tile size used in the segmentation process. This mismatch results in insufficient representation of the grain's features during clustering, leading to inaccurate segmentation. Reducing the tile size to address this issue introduces another challenge—smaller tile sizes reduce spatial resolution, which degrades the quality of the sinogram and contributes to noise during feature extraction. These observations emphasize the trade-offs involved in selecting tile size and underscore the need for careful parameter tuning to balance segmentation accuracy with spatial resolution.

Upon benchmarking, the overall workflow demonstrates high throughput processing of SEM images of dimensions 900x1000 pixels at a rate of approximately 4 images per minute on an Intel i9-12900H CPU paired with 16GB of DDR5 RAM. This efficiency combined with the unsupervised nature of workflow and its ability to postprocess and fine tune output, large-scale analysis becomes feasible, making the workflow well-suited for handling extensive datasets.

**D. Enabled Science Linking Processing Conditions to Grain Size Distributions**

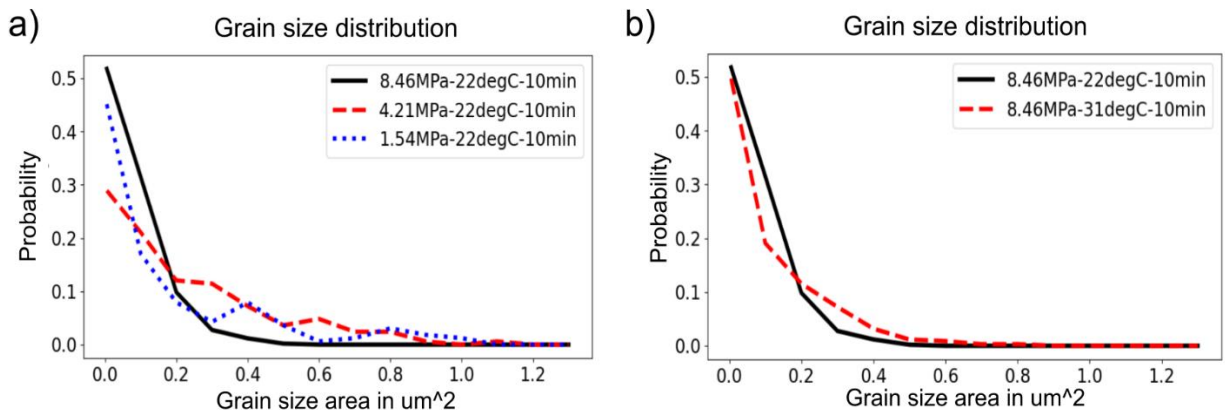

*Figure 7. Grain size distribution obtained from grains segmented with the optimal workflow for conditions where, (a) temperature is constant and pressure varies; and (b) pressure is constant and temperature varies.*

We applied the optimal workflow to identify grains and grain size distribution from SEM images obtained at varying temperature and pressure during sintering. The resulting grain index maps are analyzed to measure the areas of the segmented grains resulting in grain area distributions across multiple SEM images from each sample. We calculate grain size distributions across multiple SEM images for each sample, to understand how grain sizes are influenced by varying experimental conditions. Specifically, we examine the effect of temperature on grain size while holding pressure constant as well as examine the effect of pressure while maintaining a constant temperature. The results, as shown in **Figure 7**, give a visual understating of the grain size trends under different conditions.

For samples at constant temperature (22°C) but varying pressure, the average grain size exhibits a non-monotonic trend with increasing pressure. At 1.54 MPa, the average grain area is 0.2206 µm², which is comparable to 0.2465 µm² observed at 4.21 MPa. However, at a higher pressure of 8.46 MPa, we observe a significant reduction in grain size, with an average area of 0.108 µm². Conversely, for samples maintained at a constant pressure of 8.46 MPa but varying temperatures, the average grain size shows an increasing trend with temperature. At 22°C, the grain area averages 0.108 µm², while at 31°C, it increases to 0.1315 µm². These findings suggest a sensitivity of grain size to both pressure and temperature. However, the limited number of experimental data points restricts our ability to draw definitive conclusions regarding the trends and their underlying mechanisms. Further experiments with additional samples and varied conditions are needed to establish more robust insights into the relationship between experimental parameters and grain size distributions.

## V. Conclusion

Grain segmentation in SEM images provides critical insights into material properties, but manual analysis of such images is labor-intensive and impractical. The proposed computational workflow for segmentation of grains in SEM images leverages Radon transforms and agglomerative hierarchical clustering in an unsupervised framework, eliminating the need for manually annotated data and significantly reducing manual intervention. Through systematic testing and visual analysis, we demonstrate the workflow's capability to accurately identify and segment grains in SEM images from nanoparticle superlattices. Our results indicate that the workflow achieves high accuracy in grain segmentation while maintaining robustness against challenging edge cases, such as noisy images or regions with overlapping grain patterns. Furthermore, we demonstrate the workflows' ability to be applied to analysis of grain sizes in the experimental dataset. The workflow's processing speed, handling SEM images at a rate of four per minute on standard computational hardware, underscores its suitability for large-scale analyses.

**Open-source Code availability**
Python code and interactive Jupiter notebook available at,
https://github.com/arthijayaraman-lab/MURI_additive_SAXS_SEM/tree/main/grain_segmentation

**Acknowledgements**
Authors are grateful for financial support from the Multi University Research Initiative (MURI) from the Office of Naval Research (ONR) award # N00014-23-1-2499.

**Credit Statement**

>**Aanish Paruchuri:** Conceptualization, Methodology, Software, Validation, Formal analysis, Writing - Original Draft, Writing - Review & Editing
>**Carl Thrasher**: Conceptualization, Investigation, Data Curation, Writing - Review & Editing
>**John Hart:** Funding acquisition, Resources, Project administration, Supervision
>**Robert Macfarlane:** Funding acquisition, Resources, Project administration, Supervision
>**Arthi Jayaraman**: Funding acquisition, Resources, Project administration, Supervision, Writing - Original Draft, Writing - Review & Editing

**References**

[1] Z. Nie, A. Petukhova, and E. Kumacheva, "Properties and emerging applications of self-assembled structures made from inorganic nanoparticles," *Nat Nanotechnol*, vol. 5, no. 1, pp. 15–25, 2010, doi: 10.1038/nnano.2009.453.

[2] C. L. Bassani *et al.*, "Nanocrystal Assemblies: Current Advances and Open Problems," *ACS Nano*, vol. 18, no. 23, pp. 14791–14840, Jun. 2024, doi: 10.1021/acsnano.3c10201.

[3] M. S. Lee, D. W. Yee, M. Ye, and R. J. Macfarlane, "Nanoparticle Assembly as a Materials Development Tool," *J Am Chem Soc*, vol. 144, no. 8, pp. 3330–3346, Mar. 2022, doi: 10.1021/jacs.1c12335.

[4] Z. Cheng *et al.*, "Tuning Lattice Strain of Copper Particles in Cu/ZnO/Al2O3 Catalysts for Methanol Steam Reforming," *Energy & Fuels*, vol. 38, no. 16, pp. 15611–15621, Aug. 2024, doi: 10.1021/acs.energyfuels.4c02617.

[5] R. L. Li, C. J. Thrasher, T. Hueckel, and R. J. Macfarlane, "Hierarchically Structured Nanocomposites via a 'Systems Materials Science' Approach," *Acc Mater Res*, vol. 3, no. 12, pp. 1248–1259, Dec. 2022, doi: 10.1021/accountsmr.2c00153.

[6] P. J. Santos, P. A. Gabrys, L. Z. Zornberg, M. S. Lee, and R. J. Macfarlane, "Macroscopic materials assembled from nanoparticle superlattices," *Nature*, vol. 591, no. 7851, pp. 586–591, 2021, doi: 10.1038/s41586-021-03355-z.

[7] F. J. Humphreys, "Review Grain and subgrain characterisation by electron backscatter diffraction," *J Mater Sci*, vol. 36, no. 16, pp. 3833–3854, 2001, doi: 10.1023/A:1017973432592.

[8] R. V Thaner *et al.*, "The Significance of Multivalent Bonding Motifs and 'Bond Order' in DNA-Directed Nanoparticle Crystallization," *J Am Chem Soc,* vol. 138, no. 19, pp. 6119–6122, May 2016, doi: 10.1021/jacs.6b02479.

[9] S. E. Seo, M. Girard, M. O. de la Cruz, and C. A. Mirkin, "The Importance of Salt-Enhanced Electrostatic Repulsion in Colloidal Crystal Engineering with DNA," *ACS Cent Sci*, vol. 5, no. 1, pp. 186–191, Jan. 2019, doi: 10.1021/acscentsci.8b00826.

[10] T. B. Gaines, C. Boyle, J. M. Keller, M. R. Maschmann, S. Price, and G. J. Scott, "Scanning Electron Microscope Image Segmentation with Foundation AI Vision Model for Nanoparticles in Autonomous Materials Explorations," in *2024 IEEE Conference on Artificial Intelligence (CAI)*, 2024, pp. 1266–1271. doi: 10.1109/CAI59869.2024.00224.

[11] A. Shah *et al.*, "Automated image segmentation of scanning electron microscopy images of graphene using U-Net Neural Network," *Mater Today Commun*, vol. 35, p. 106127, 2023, doi: https://doi.org/10.1016/j.mtcomm.2023.106127.

[12] K. He, G. Gkioxari, P. Dollár, and R. Girshick, "Mask r-cnn," in *Proceedings of the IEEE international conference on computer vision*, 2017, pp. 2961–2969.

[13] O. Ronneberger, P. Fischer, and T. Brox, "U-net: Convolutional networks for biomedical image segmentation," in *Medical image computing and computer-assisted intervention–MICCAI 2015: 18th international conference, Munich, Germany, October 5-9, 2015, proceedings, part III 18*, Springer, 2015, pp. 234–241.

[14] K. P. Treder, C. Huang, J. S. Kim, and A. I. Kirkland, "Applications of deep learning in electron microscopy," *Microscopy*, vol. 71, no. Supplement_1, pp. i100–i115, 2022.

[15] L. Zhang and S. Shao, "Image-based machine learning for materials science," *J Appl Phys*, vol. 132, no. 10, 2022.

[16] S. Lu, X. Zhao, H. Liu, and H. Liang, "Semiconductor Material Porosity Segmentation in Flame Retardant Materials SEM Images Using Data Augmentation and Transfer Learning," in *Advances in Transdisciplinary Engineering*, IOS Press BV, Feb. 2024, pp. 74–81. doi: 10.3233/ATDE240011.

[17] S. Lu and A. Jayaraman, "Pair-Variational Autoencoders for Linking and Cross-Reconstruction of Characterization Data from Complementary Structural Characterization Techniques," *JACS Au*, vol. 3, no. 9, pp. 2510–2521, Sep. 2023, doi: 10.1021/jacsau.3c00275.

[18] J. Stuckner, B. Harder, and T. M. Smith, "Microstructure segmentation with deep learning encoders pre-trained on a large microscopy dataset," *NPJ Comput Mater*, vol. 8, no. 1, p. 200, 2022, doi: 10.1038/s41524-022-00878-5.

[19] A. Paruchuri, Y. Wang, X. Gu, and A. Jayaraman, "Machine learning for analyzing atomic force microscopy (AFM) images generated from polymer blends," *Digital Discovery*, 2024.

[20] J. Zhang, P. J. Santos, P. A. Gabrys, S. Lee, C. Liu, and R. J. Macfarlane, "Self-Assembling Nanocomposite Tectons," *J Am Chem Soc*, vol. 138, no. 50, pp. 16228–16231, Dec. 2016, doi: 10.1021/jacs.6b11052.

[21] P. J. Santos, T. C. Cheung, and R. J. Macfarlane, "Assembling Ordered Crystals with Disperse Building Blocks," *Nano Lett*, vol. 19, no. 8, pp. 5774–5780, Aug. 2019, doi: 10.1021/acs.nanolett.9b02508.

[22] P. J. Santos, Z. Cao, J. Zhang, A. Alexander-Katz, and R. J. Macfarlane, "Dictating Nanoparticle Assembly via Systems-Level Control of Molecular Multivalency," *J Am Chem Soc*, vol. 141, no. 37, pp. 14624–14632, Sep. 2019, doi: 10.1021/jacs.9b04999.

[23] C. J. Thrasher *et al.*, "Rationally Designing the Supramolecular Interfaces of Nanoparticle Superlattices with Multivalent Polymers," *J Am Chem Soc*, vol. 146, no. 16, pp. 11532–11541, Apr. 2024, doi: 10.1021/jacs.4c02617.

# SUPPORTING INFORMATION

# Machine Learning for Identifying Grain Boundaries in Scanning Electron Microscopy (SEM) Images of Nanoparticle Superlattices – Supporting information

Aanish Paruchuri [1], Carl Thrasher[2], A. J. Hart[3], Robert Macfarlane[2] Arthi Jayaraman [4,5,6] *

1. Master of Science in Data Science Program, University of Delaware, Newark DE 19713
2. Department of Materials Science and Engineering, Massachusetts Institute of Technology, Cambridge, MA, 02139
3. Department of Mechanical Engineering, Massachusetts Institute of Technology, Cambridge, MA, 02139
4. Department of Chemical and Biomolecular Engineering, 150 Academy St, University of Delaware, Newark DE 19713
5. Department of Materials Science and Engineering, University of Delaware, Newark DE 19713
6. Data Science Institute, University of Delaware, Newark DE, 19713

**\* Corresponding author**

arthij@udel.edu

*Figure S1*: *(a-d) Examples of SEM images in the dataset along with instruments metadata represented in the bottom pane of the image. (a-d) shows varying contrast and extent of defects along with line noise observed in (b, d).*

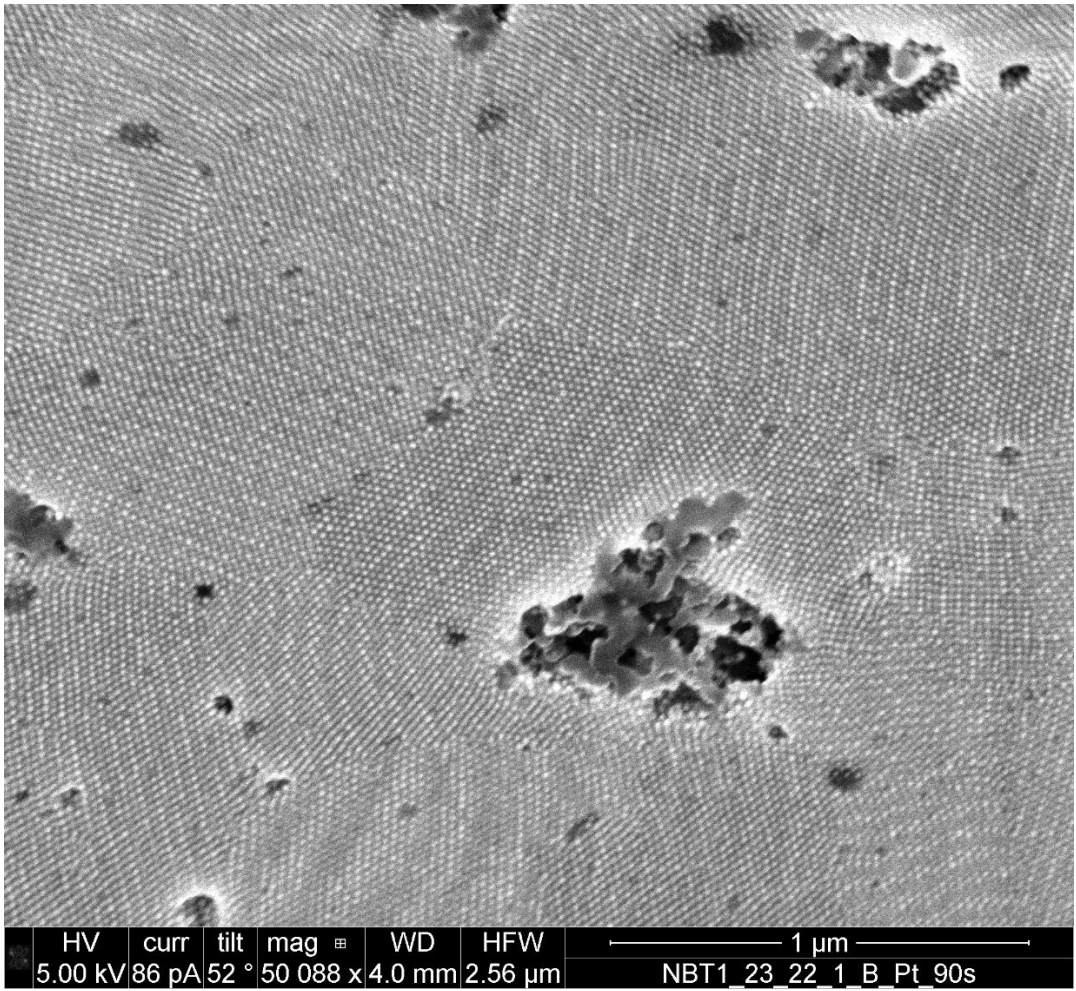

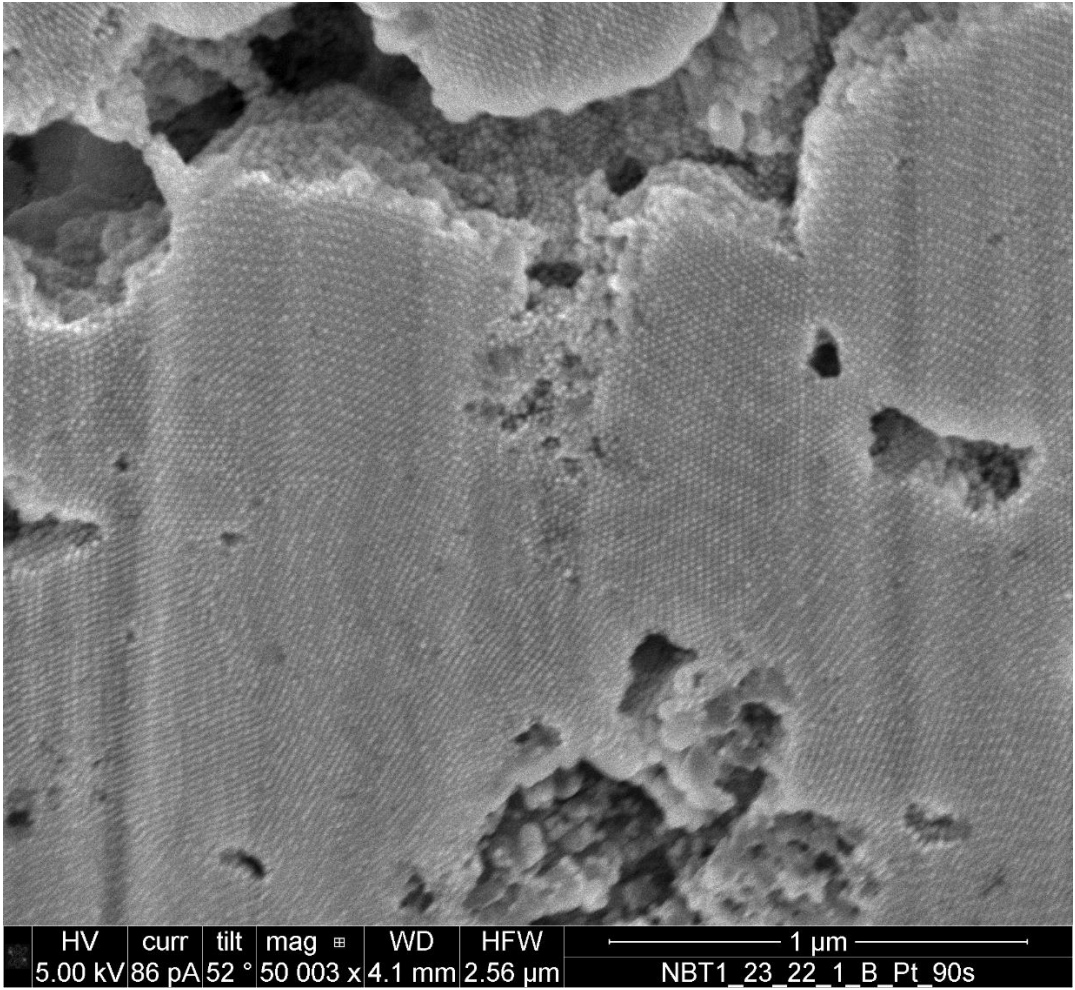

c

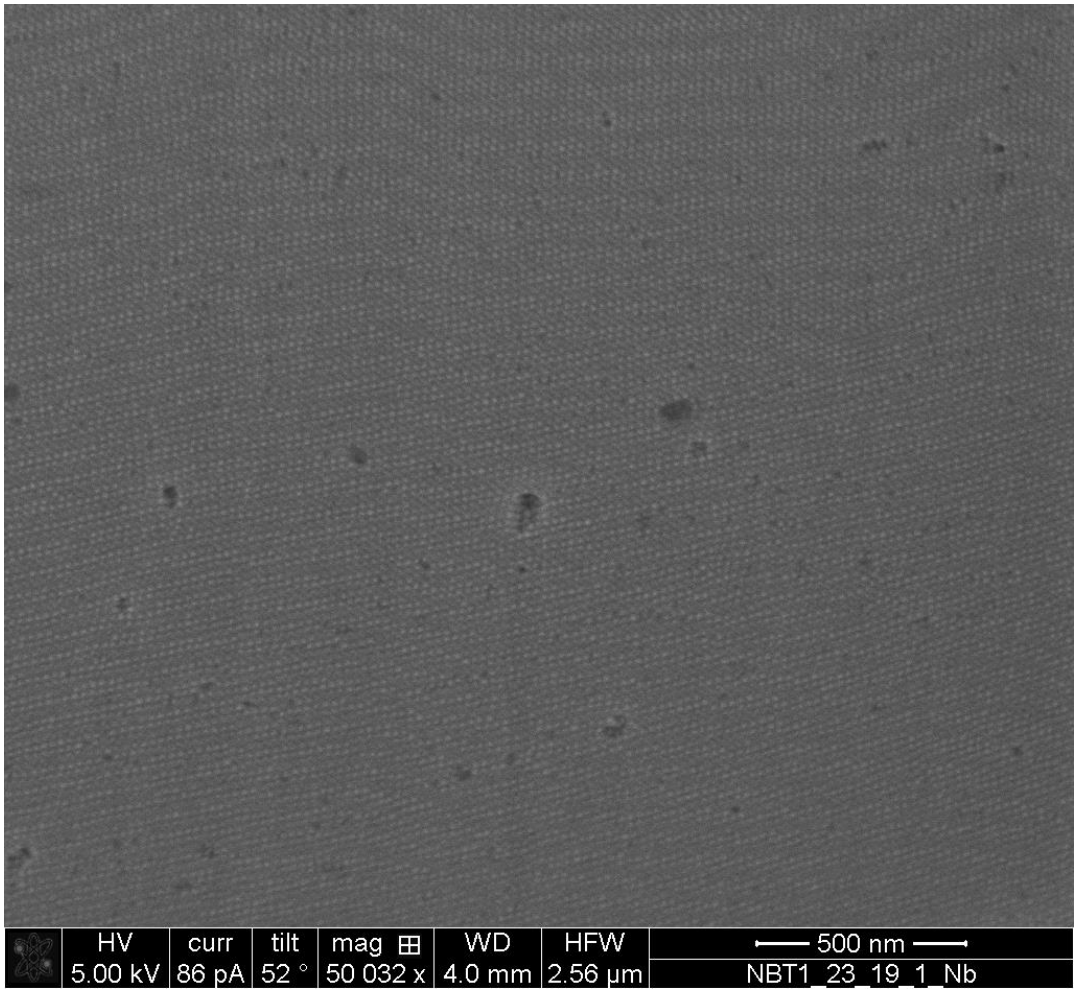

d

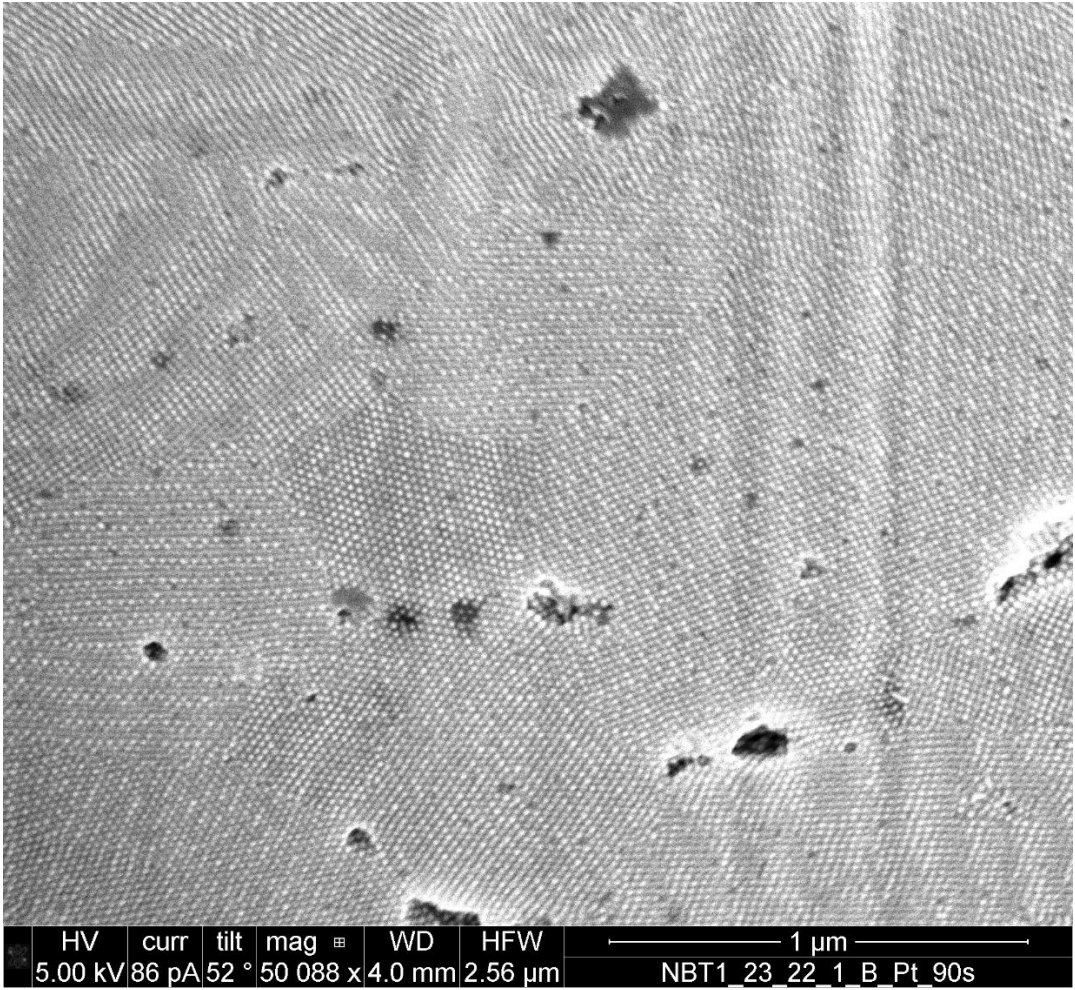

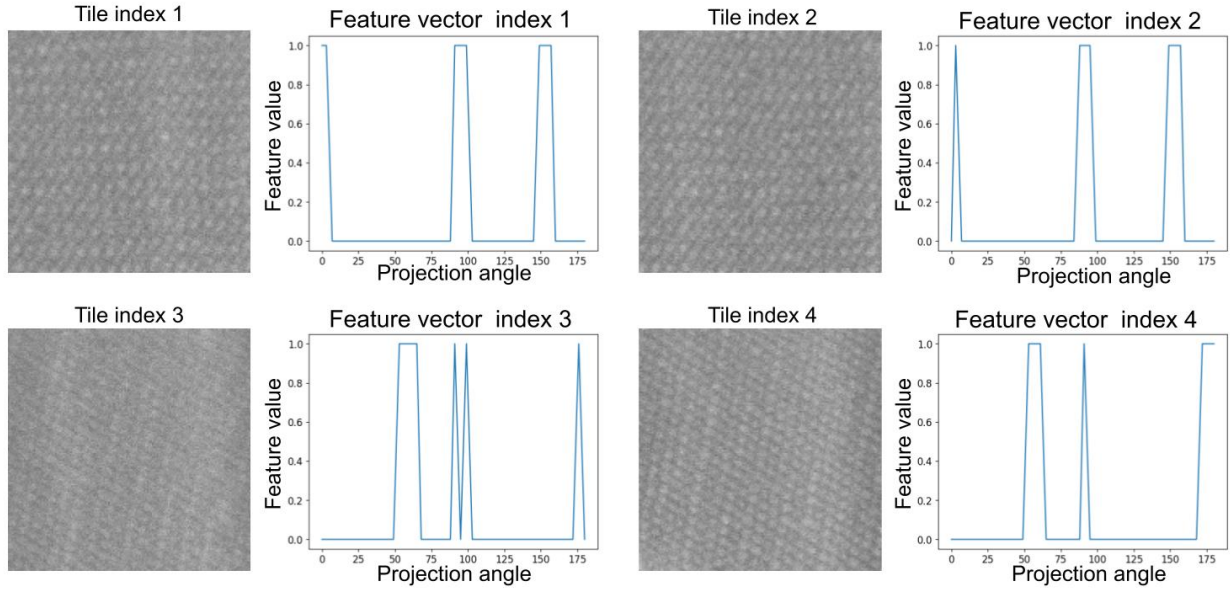

*Figure S2*: *Feature vectors are extracted using the Radon transform applied to individual tiles. The Dynamic Time Warping (DTW) distance between tiles with similar grains is typically less than 1.41, while tiles with differing grains also exhibit DTW distances below this threshold. This overlap causes tiles with distinct grains to be misclassified as similar, leading to segmentation errors and noisy predictions.*